% TEX SET-UP FOR PROCEEDINGS OF "ASI MEETING HELD AT AHMEDABAD"
% TO BE PUBLISHED IN THE BULLETIN OF ASTRONOMICAL SOCIETY OF INDIA
% Please do not change the page layout, except \hoffset and/or
% \voffset, where changes may be required depending on default printer
% settings. Leave a margin of 3.8 cm on the left and 5 cm at the top
\input psfig.sty
\magnification=\magstep0
\hsize=13.5 cm               %  horizontal size of printed page
\vsize=19.0 cm               %  vertical size of printed page
\baselineskip=12 pt plus 1 pt minus 1 pt  % The line spacing
\parindent=0.5 cm  % The paragraph indent
\hoffset=1.3 cm      % The horizontal offset (may need to be changed)
\voffset=2.5 cm      % The vertical offset (may need to be changed)
\font\twelvebf=cmbx10 at 12truept % Set bold font for Title
\font\twelverm=cmr10 at 12truept % Set large font for Name
\overfullrule=0pt
\nopagenumbers    %  Actual page nos. will be inserted by the Editor
%
% The headlines
% The changes in the headlines should be made just before the Abstract
\newtoks\leftheadline \leftheadline={\hfill {\eightit S. K. Saha and 
V. Chinnappan}
\hfill}
\newtoks\rightheadline \rightheadline={\hfill {\eightit Performance of the low 
light level CCD camera for speckle imaging}
 \hfill}
% Do not change the headline on the first page of paper.
\newtoks\firstheadline \firstheadline={{\eightrm Bull. Astron. Soc.
India (2002) {\eightbf 30,} } \hfill}
\def\makeheadline{\vbox to 0pt{\vskip -22.5pt
\line{\vbox to 8.5 pt{}\ifnum\pageno=1\the\firstheadline\else%
\ifodd\pageno\the\rightheadline\else%
\the\leftheadline\fi\fi}\vss}\nointerlineskip}
%
% Defining 8-pt fonts for figure captions and references
\font\eightrm=cmr8  \font\eighti=cmmi8  \font\eightsy=cmsy8
\font\eightbf=cmbx8 \font\eighttt=cmtt8 \font\eightit=cmti8
\font\eightsl=cmsl8
\font\sixrm=cmr6    \font\sixi=cmmi6    \font\sixsy=cmsy6
\font\sixbf=cmbx6
%
%for switching to eight point type \eightpoint
\def\eightpoint{\def\rm{\fam0\eightrm}
\textfont0=\eightrm \scriptfont0=\sixrm \scriptscriptfont0=\fiverm
\textfont1=\eighti  \scriptfont1=\sixi  \scriptscriptfont1=\fivei
\textfont2=\eightsy \scriptfont2=\sixsy \scriptscriptfont2=\fivesy
\textfont3=\tenex   \scriptfont3=\tenex \scriptscriptfont3=\tenex
\textfont\itfam=\eightit  \def\it{\fam\itfam\eightit}%
\textfont\slfam=\eightsl  \def\sl{\fam\slfam\eightsl}%
\textfont\ttfam=\eighttt  \def\tt{\fam\ttfam\eighttt}%
\textfont\bffam=\eightbf  \scriptfont\bffam=\sixbf
\scriptscriptfont\bffam=\fivebf \def\bf{\fam\bffam\eightbf}%
\normalbaselineskip=10pt plus 0.1 pt minus 0.1 pt
\normalbaselines
\abovedisplayskip=10pt plus 2.4pt minus 7pt
\belowdisplayskip=10pt plus 2.4pt minus 7pt
\belowdisplayshortskip=5.6pt plus 2.4pt minus 3.2pt \rm}
%
% define the displayed equations to be indented 1.5 cm from left
% as required by the Bulletin. Hopefully this will work for all
% equations. With this definition using the normal $$...$$ should
% produce the equations with correct indentation, but it will be necessary
% to use \eqno to put equation numbers (though it will be possible to put
% blank eq. nos.). Further, eq. nos using \eqalignno will not work
%
\def\leftdisplay#1\eqno#2$${\line{\indent\indent\indent%
$\displaystyle{#1}$\hfil #2}$$}
\everydisplay{\leftdisplay}
%
% Some useful definitions
% less than or order of \la
\def\frac#1#2{{#1\over#2}}

% greater than or order of \ga

%
%
%to generate boldface characters
\def\pmb#1{\setbox0=\hbox{$#1$}\kern-0.015em\copy0\kern-\wd0%
\kern0.03em\copy0\kern-\wd0\kern-0.015em\raise0.03em\box0}
%
%Beginning of Document%
\pageno=1
\vglue 50 pt  %Leave some space on page 1 before the title
% The title
%
\leftline{\twelvebf Performance of the low light level CCD camera for speckle 
imaging}
% if more than one line is required for the title, then use next two lines ...
%
\smallskip
%\leftline{\twelvebf Proceedings of ``ASI meeting held at Ahmedabad''}
% end of title
\vskip 40 pt  % Space between title and author(s) name(s).
\leftline{\twelverm S. K. Saha\footnote{$^*$}{\eightit e-mail: 
sks@iiap.ernet.in, vchin@iiap.ernet.in} and V. Chinnappan$^*$} 
% Name of Authors
\vskip 4 pt
\leftline{\eightit Indian Institute of Astrophysics, Bangalore 560 034, India.
}
%\leftline{\eightit  to reduce the number of lines}
%
% If authors are from different institutes, repeat the above lines
% for each institution. For authors from same institution write the
% names in one line.
%
%\vskip 0.5 cm
%\leftline{\twelverm V. R. Co-author1 and V. R. Co-author2}
%\vskip 4 pt
%\leftline{\eightit Name and Address of the institution}
\vskip 20 pt % leave some space between author(s) names(s) and abstract
%
%
% The leftheadline should include the Authors' name, for two authors use
% \&  (e.g. I. M. Author \& I. M. Co-author) for three or more authors
% use et al.,
\leftheadline={\hfill {\eightit S. K. Saha and V. Chinnappan} \hfill}
% Use a short running title as the rightheadline
\rightheadline={\hfill {\eightit Performance of the low light level CCD camera 
for speckle imaging} \hfill}

% Abstract begins
%
{\parindent=0cm\leftskip=1.5 cm

{\bf Abstract.}
\noindent
A new generation CCD detector called low light level CCD (L3CCD) that performs 
like an intensified CCD without incorporating a micro channel plate (MCP) for 
light amplification was procured and tested. A series of short exposure images 
with millisecond integration time has been obtained. The L3CCD is cooled to 
about $-80^\circ$~C by Peltier cooling. 
\smallskip 
\vskip 0.5 cm  %  Space between Abstract and Key words
{\it Key words:} interferometer, speckle imaging, L3CCD camera.

}                                 %  End of abstract
% Beginning of document
%
%
% Beginning of a section heading
%
% for the first section leave 20 pt space, for subsequent sections just
% leave bigskip (i.e. 12 pt)
\vskip 20 pt
\centerline{\bf 1. Introduction}
\bigskip
\noindent  
High resolution imagery requires a high quality sensor that helps in obtaining 
snap shots of very high time resolution, depending on the dwell time of the 
speckle pattern, of the order of either (i) frame integration of 50~Hz, or 
(ii) photon recording rates of a few MHz. Recent
development of a non-intensified low light level charge coupled device (L3CCD) 
which effectively the reduces readout noise to less than one electron rms has 
enabled substantial internal gain within the CCD before the signal reaches the
output amplifier (Mackay et al. 2001). Such a detector shows
promise for quantitative measurement of diffraction-limited stellar images
with suitable instruments, such as a speckle interferometer. We have procured 
one such system from the Andor technology recently for recording
specklegrams of a few interesting objects, viz., close binary stars, active 
galactic nuclei (AGN) etc., at the 2.34 meter Vainu Bappu Telescope using the 
speckle interferometer developed by Saha et al. (1999).  
\vskip 20 pt
\centerline{\bf 2. Salient features of L3CCD}
\bigskip
\noindent  
In general, an intensified CCD (ICCD) detector consists of an image
intensifier (MCP) coupled to a CCD camera. The major disadvantages of ICCD
is the high voltage operation which affects its usages at high humidity and its
susceptibility to damage when more light is allowed to fall on to the 
intensifier. Another drawback of such system is the poor gain statistics 
results in the introduction of a noise factor between 2 and 3.5.
\bigskip
\noindent  
In the L3CCD camera, no safety measures are required under
bright light. The system is provided with Peltier cooling that
operates to - 65$^\circ$C with air-cooling and with further additional water
circulation, it reaches -80$^\circ$C. The performance of this cooling
system is comparable with a liquid nitrogen cooled cryostats. The L3CCD is a 
frame transfer device where the image store and readout register are of
conventional design that operates typically at 10 volts. But there is an 
extended section of gain register between the normal serial register and the 
final detection node which operates at much higher amplitude (typically
at 40-50 volts). This large voltage creates an avalanche multiplication
which thereby increases the number of electrons in the charge packets, thus
producing a gain. Adjustment of the gain is possible with fine control of the
voltage. All the output signals above a threshold may be counted as photon 
events provided the incoming photon flux is of a sufficiently low intensity that
no more than one electron is generated in any pixel during the integration 
period, and the dark noise is zero, and the gain is set at a suitable level 
with respect to the amplifier read noise. 
\bigskip
\noindent  
The L3CCD camera system that we have procured has 576$\times$288 pixels of size
20$\times$30$\mu$m$^2$ in the image area. The storage section contains 591 
columns and 288 rows. The physical area of the chip is $11.52 \times 8.64$~mm. 
It has the provision to change the gain from 1 to 1000 by software.
The noise at 1~MHz read rate is 0.1 e. Each pixel data is digitized to 16
bit resolution. The system has full vertical binning mode
suitable for spectroscopy, single scan for regular astronomical
imaging with long integration time, kinetic series acquisition, and the
frame transfer mode with region of interest that can have short integration 
time in the order of milliseconds. Several hundreds of frames in a series can 
be recorded in this mode. For speckle observations, the kinetic mode acquisition
mode was used. 
\vskip 20 pt
\centerline{\bf 3. Performance} 
\bigskip 
\noindent
We have tested this L3CCD both at the laboratory, as well as at the 2.34~m
VBT, VBO, Kavalur. The performance was found to be satisfactory. The dark count 
was found to be 0.001 e/pixel/second at -80$^\circ$ and the saturation signal 
was about 52, 000 counts/pixel. Without external power supply and using the 
internal PC power, the Peltier cooler reached the set temperature of 
$-50^\circ~C$ from the ambient temperature of $20^\circ~C$ within 11 minutes.
\bigskip
\noindent
On 15th December, 2001, we have recorded an image of the star, HD36151 
(m$_v$ 6.5) at the Cassegrain focus of 2.34 m VBT, VBO, Kavalur, with an 
exposure time of 10 msec, with the software gain set to 150, and a CCD 
sensitivity of 6.14 e per A/D count. The recorded image shows 50,000 counts. 
It may be noted here that
the software gain can be varied from 1 to 255, which nonlinearly corresponds 
to an actual gain of 1 to 1000. A 12th magnitude star with an integration time 
of 10 msec. has been recorded as well. The count was found to be about 2000; 
software gain was set to 200.
\vskip 20 pt
\centerline{\bf 3. Discussion}
\bigskip 
\noindent
For experiments like speckle imaging (Labeyrie, 1970) and adaptive optics where 
the integration time is dictated by the atmospheric coherence time, which is 
normally of the order of a few milliseconds, this L3CCD may be more suitable 
than the MCP-based detectors. For the selective image reconstruction 
technique (Dantowitz et al. 2000), where a few sharpest images are selected from
a large dataset of short-exposures images,
L3CCD detector will provide important advantages (Baldwin et al. 2001). 
\bigskip
\centerline{\bf References}
\bigskip
{\eightpoint\parindent=0pt\everypar={\hangindent=0.5 cm}
% References in the format of the Bulletin of the Astronomical Society of India
% using 8 pt fonts
% leave one line blank between two references to force a paragraph break
%scussions.

Baldwin J. E., Tubbs R. N., Cox G. C., Mackay C. D., Wilson R. W., and Anderson
M. I., 2001, A \& A., 368, L1.

Dantowitz R. F., Teare S., and Kozubal M., 2000, A J, 119, 2455.

Labeyrie A., 1970, A \& A., 6, 85.

Mackay C. D., Tubbs R. N., Bell R., Burt R., and Moody I., 2000, SPIE, 4306.

Saha S. K., Sudheendra G., Umesh Chandra A., Chinnappan V., 1999, 
Experimental Astronomy.  9, 39.  

% End of section heading
%\endref
}                                         % End of references
% leave one line blank before the closing braces.
\end